\documentclass{aa}
\usepackage{graphicx}

\begin{document}
 
\title{Fast growth of supermassive black Holes in galaxies}

\author{Faustin Munyaneza\thanks{Humboldt Fellow} \and Peter L. Biermann}
\offprints{ F. Munyaneza,
\email{munyanez@mpifr-bonn.mpg.de}}
\institute{Max-Planck-Institut  f\"ur Radioastronomie,
         Auf Dem H\"ugel 69,
         D-53121 Bonn, 
         Germany}
\date{Received <date>/ Accepted <date>}

\abstract{
We report on  a calculation of the growth of the mass of  
supermassive black holes  at galactic centers from dark matter and 
Eddington - limited baryonic accretion. 
Assuming that
dark matter halos are made of fermions and harbor 
 compact degenerate Fermi balls 
of masses from 
$10^{3}M_{\odot}$ to $10^{6}M_{\odot}$,
we find that dark matter accretion can boost the mass of 
seed black holes from about $\sim 5M_{\odot}$ to 
$10^{3-4}M_{\odot}$ black holes, which then 
grow by Eddington - limited baryonic accretion to 
supermassive black holes of $10^{6 \, - \, 9}M_{\odot}$. 
We then  show that the formation of the recently 
detected supermassive black hole
of $3\times 10^{9}M_{\odot}$  at a 
redshift of $z = 6.41$ in the quasar SDSS J114816.64+525150.3
could be understood if the black hole   completely consumes  the 
degenerate Fermi ball
and then grows by Eddington - limited baryonic accretion.
In the context of this model we constrain the dark matter particle masses 
to be within the 
range from 12 ${\rm keV/c}^{2}$ to about 450 ${\rm keV/c}^{2}$.
Finally we investigate the black hole  growth dependence 
 on the formation time of the seed BH
and on the mass of the seed BH. We find that in order to
fit the observed data point of $M_{BH} \sim 3 \times 10^{9}M_{\odot}$
and $z \sim 6.41$, dark matter accretion cannot start later
than about $2 \times 10^{8}$ years and the seed BH cannot be
greater than about $10^{4}M_{\odot}$.
 Our results are in full agreement with the 
WMAP observations that indicate that the first onset of 
star formation might have occurred at a 
redshift of $z \sim 15 \; - \; 20$. For other models of dark matter particle masses,
corresponding constraints may be derived from the growth of black holes in the 
center of galaxies.

}

\titlerunning{Growth of  black holes}
\authorrunning{F. Munyaneza \& P.L. Biermann}
\maketitle

\section{Introduction}
Over the past few years, the idea of dark matter (DM) and the 
possible  existence of  supermassive black holes (BH) of masses 
from $10^{6.5}$ to $10^{9.5}M_{\odot}$ at the center of 
galaxies (Macchetto et al. \cite{macchetto97}; Sch\"odel et al. \cite{schodel02}) have  become deeply rooted within the astrophysical community. The exploration of the relationship between these two intriguing problems of modern astrophysics has been the subject of an increasing number of papers.

It has been established  that the mass of the central BH is tightly correlated with the velocity dispersion $\sigma$ of its host bulge, where it is found that $M_{BH} \sim
\sigma^{4-5}$ (Faber et al. \cite{faber97}; Magorrian et al. \cite{magorian98};
Ferrarese \& Merritt \cite{ferrase00}; Gebhardt et al. \cite{gebhardt00};  Ferrarese \cite{ferrase02}; Haering \& Rix \cite{rix04}). This tight relation between  the masses of the BHs and the gravitational potential well that hosts them
 suggests that the formation and evolution of supermassive BHs and the bulge of the 
parent galaxy may be closely
related, e.g.  Wang, Biermann \& Wandel (\cite{wang00}).
In addition,  the recent discovery of high redshift quasars 
with $z > 6 $ (Fan et al. \cite{fan01})
implies that the formation  of supermassive BHs 
took place over fewer than $10^{9}$years. In spite of the vast and tantalizing work undertaken on BHs,
  their  genesis and evolution  are not well understood (see Rees \cite{rees84} for a review).
Since the discovery of quasars in the early 1960s, it has 
been suggested that these objects are powered by accretion of gas onto the supermassive BHs of masses $10^{6}-10^{9}M_{\odot}$ (Lynden-Bell \cite{bell69}). Two scenarios have been discussed in modeling the growth of BHs. One is that BHs grow out of a low mass `seed' BH through accretion (Rees \cite{rees84}), and another one is that BHs grow by  
merging (Barkana et al. \cite{barkana01}; Wang, Biermann \& Wandel \cite{wang00}; 
Gopal-Krishna, Biermann \& Wiita \cite{bk03,bk04}). In a recent paper, Duschl \&
Strittmatter (\cite{ds04}) have investigated a model for the 
formation of supermassive black
holes using a combination of merging and accretion mechanisms.

The purpose  of this paper is to study
 the growth of BHs from dark matter and Eddington-limited baryonic accretion.
 Also, we would like to obtain some constraints on the DM particle masses. 
In the past, self gravitating neutrino matter
has been suggested  as a model
for quasars, with neutrino masses in the 
range $0.2 {\rm keV} \stackrel {\textstyle
<} {\sim} m_{f}   \stackrel {\textstyle
<} {\sim} 0.5 {\rm MeV}$ (Markov \cite{markov64}).
Later, neutrino matter was suggested to describe DM in clusters of galaxies
and galactic halos with masses in the range of
$1 {\rm eV} \stackrel {\textstyle
<}{\sim} m_{f}   \stackrel{\textstyle
<}{\sim} 25 {\rm eV}$ (Cowsick \& McClelland \cite{cowsik73}; Ruffini
\cite{ruffini80}).
More recently, fermion balls (FBs) made of degenerate fermionic matter
of  $10 \ {\rm keV} \stackrel {\textstyle
<}{\sim} m_{f}   \stackrel{\textstyle
<} {\sim} 25 \ {\rm keV}$ were
suggested as an alternative to supermassive BHs in galaxies
(Viollier \cite{viollier94}; Bili\'c, Munyaneza \& Viollier \cite{bmv99};
Tsiklauri \& Viollier \cite{tv98}; Munyaneza \& Viollier \cite{mv02}).
It has been also suggested that if the Galaxy harbors a supermassive BH, then
there should be a density spike in which dark matter (DM) falling towards the
center could annihilate and the detection of these annihilation signals
could be used as a probe for the nature of DM (Bertone, Silk \& Sigl
\cite{bertone02}; Gondolo \& Silk \cite{gondolo99} and  Merritt et al.
\cite{merritt02}).
The current belief is that DM particles are bosonic and very massive, i.e.
$m_{DM} \stackrel {\textstyle
>}{\sim} 1 GeV/c^{2}$. 
However, the absence of experimental constraints on the weakly interacting
massive particles (WIMPs) that probably constitute DM leaves the door 
open for further investigation of the hidden mass of the Universe.

In this paper, we will assume DM to be of fermionic matter and described by
a Fermi - Dirac distribution with an energy  cutoff in phase space (King 
\cite{king66}). We will then explore the limits for the DM particle masses
 in
order to reproduce the mass distribution in the Galaxy (Wilkinson \& Evans \cite{evans99}) and then study the growth of a seed BH immersed at the
center of the DM distribution in galaxies.
We use degenerate FBs at the center of DM halos
 not as 
replacements  for 
the BHs but as necessary ingredients to grow the BHs in galactic centers.
 
The resulting distribution of stars around a massive BH was studied in 
detail in the  1970s and early 1980s in the context of 
globular clusters (Hills \cite{hills75}; Frank \& Rees \cite{frank76};
Bahcall \& Wolf \cite{bahcal76};  Duncan \& Shapiro \cite{duncan82};
 Shapiro  \cite{shapiro85}). Peebles (\cite{peeble72})  studied the adiabatic 
 growth of a BH in an isothermal sphere and showed that  the BH would alter the 
 matter density to  an adiabatic cusp with $\rho \sim r^{-3/2}$. A few years later,
   Young (\cite{young80}) constructed numerical models that  confirmed Peebles' results and showed that the BH induces a tangential anisotropy in the velocity dispersion.
In this paper, we use the results of previous calculations on  the 
growth of BHs that are accreting stars
(Lightman \& Shapiro \cite{lightman78}) to  investigate the 
growth of a BH that  accretes DM.
 Here we assume that the physics driving the formation of the power law cusp
   in the star - star case is the same as in the case of DM particle
  orbits being perturbed by molecular clouds.  Julian (1967)
   investigated a similar scenario in which the stellar orbits in our Galaxy were perturbed by
  molecular clouds to explain the stellar velocity dispersion  dependence on the star's age.
  Moreover, Duncan \& Wheeler (\cite{dw80}) investigated the anisotropy of the velocity dispersions of the stars
  around the BH in M87. 

  
Given the density distribution in DM halos, we are interested in 
establishing how a seed BH would grow by accreting DM.
  Moreover, the comparison of the growth of the BH from accretion of DM and 
  Eddington - limited baryonic matter would give us
  another piece of information in the debate surrounding the nature of DM.
 We therefore investigate how a BH 
seed of $5 \  M_{\odot}$ could grow 
to a $3 \times 10^{9}M_{\odot}$ BH as 
recently detected in  quasar  SDSS J1148+5251 at z=6.41 (Willot, McLure \& Jarvis \cite{willot03}). Such a seed BH of a typical mass between 5 and 9 $M_{\odot}$ could have evolved in BH binaries (Podsiadlowski, Rappaport \& Hau \cite{pod03}).
  Here, we note that the seed BH could  be  an intermediate mass black  hole
 (IMBH)
 of $10^{3-4}M_{\odot}$ that might have formed from collisions in  dense
 star-forming regions (Portegies Zwart \& McMillan \cite{porte02}, Coleman Miller
  \cite{miller03} and van der Marel \cite{van03} for a review).
In fact,  Wang \& Biermann (\cite{wang98}) have established that a BH would 
grow exponentially with time accreting baryonic matter as long as the supply lasts.
In addition, they were able to reproduce the observed correlation $M_{BH}/M_{sph}$ using standard disk galaxy parameters.
Assuming a cusp - like distribution of self-interacting DM (Spergel \& Steinhardt \cite{spergel00}), Ostriker (\cite{ostriker00}) has estimated that BHs could grow to $10^{6}-10^{9}M_{\odot}$ from DM accretion.
For completeness, we note that accretion of DM particles by BHs has recently been 
studied in Zhao, Haehnelt \& Rees (\cite{zhao02}) 
and Read \& Gilmore (\cite{read03}).
 
Cosmological parameters of $H_{0}=70 \ {\rm km \ s^{-1} Mpc^{-1}}$, $\Omega_{m}=0.3$ and $\Omega_{\Lambda}=0.7$ are assumed throughout this paper.
 In section~2, we 
establish the main equations to describe DM in galaxies.
We then discuss the  growth of seed BHs from  DM and Eddington 
- limited baryonic matter accretion in section~3 and  conclude 
 with a discussion in section~4.

\section{Dark matter in galaxies}

\subsection{Main equations}

 We characterize DM by its mass density  $\rho(r)$ and 
its velocity dispersion $\sigma$. 
DM is assumed to be collisionless and of fermionic matter and the mass of DM particles is denoted  by $m_{f}$.
We will look for  DM distributions with  degenerate 
cores at the center of the DM halos instead of a steep power law
$\rho \sim r^{-\gamma}$ with $\gamma \approx 1-1.5$ (e.g. Navarro,
Frank \& White \cite{nfw97}) as we are assuming here that a central
cusp is a result of  BH growth.
In order to obtain bound astrophysical solutions at 
finite temperature, we follow  King (\cite{king66})  and  
introduce an energy  cutoff in phase space using the Fermi-Dirac distribution function.  
We therefore adopt the following prescription for the distribution function
\begin{eqnarray}
f(E) =
\left\{
\begin{array}{l}
 \displaystyle{\frac{g_{f}}{8\pi^{3}\hbar^{3}}\frac{1}{{\rm exp}\left(\frac{E-\mu}{kT}\right) + 1} + {\rm C} \; \; , \; \; \ E < E_{c} } \; \;  \\ [.5cm]
 \displaystyle{0 \hspace{3.35cm} , \; \; E > E_{c} } \; \; , \\
\end{array} \right.
\label{eq:02}
\end{eqnarray}
where
$g_{f}$ is the spin degree of freedom, i.e. $g_{f}=2$  for Majorana 
and $g_{f}=4$ for Dirac's fermions, $\hbar$ is the Planck constant, $E$ is the
particle energy, $\mu$ is the chemical potential 
and $T$ stands for temperature.
The constant $C$ is chosen so that the distribution function vanishes at the 
cutoff energy $E=E_{c}$. As a  result, we get the following ''lowered'' Fermi-Dirac 
 distribution that will be used throughout the paper.
\begin{eqnarray}
f(E,E_{c}) =
\left\{
\begin{array}{l}
 \displaystyle{\frac{g_{f}}{8\pi^{3}\hbar^{3}}\frac{{\rm exp}\left(\frac{E_{c}-\mu}{kT}\right)}{{\rm exp}
\left(\frac{E_{c}-\mu}{kT}\right) + 1} \times
\frac{1-{\rm exp}\left(\frac{E-E_{c}}{kT}\right)}{{\rm exp}\left(\frac{E-\mu}{kT}\right) + 1} \; \; , \; \; \ E < E_{c}  } \\ [.5cm]
 \displaystyle{0 \hspace{3.35cm} , \; \; E > E_{c}  } \; \; .\\
\end{array} \right.
\label{eq:03}
\end{eqnarray}
 The mass density $\rho(r)$ is given by
\begin{equation}
\rho(r)= m_{f} \int_{0}^{E_{c}}4 \pi f(p)p^{2}dp(E)   \, \, ,
\end{equation}
and the gravitational potential $\Phi$ obeys  Poisson's equation
\begin{equation}
\Delta \Phi(r)=4\pi G \rho(r)  \, \, .
\label{eq:05}
\end{equation}
Next we introduce a dimensionless gravitational potential $v$, a radial
variable $x$ and an inverse temperature $\beta$ as
\begin{equation}
v=\frac{r \left(\Phi_{0}-\Phi(r)\right)}{GM_{\odot}}  \, \, \, \, \, \,,
\, \, \, \,  x=\frac{r}{a} \, \, \, \, \,  , \, \, \,  \beta=\frac{T_{0}}{T} ,
\end{equation}
where $a$  and $T_{0}$ are the length and temperature scales, respectively.
They are given as
\begin{equation}
a=\left[\frac{3\pi\hbar^{3}}{2^{5/2}m_{f}^{4}g_{f}G^{3/2}M_{\odot}^{1/2}}\right]^{2/3} \; \; \; \; ,
\; \; \;  T_{0}=\frac{GM_{\odot}m_{f}}{ak} \; \;,
\end{equation}
where $k$ is  Boltzmann's constant and the cutoff 
energy $E_{c}$ has been
chosen as $E_{c}=m_{f}\Phi_{0}$. We also define the degeneracy 
parameter $\Lambda$ as
\begin{equation}
\Lambda={\rm  exp} \left( \frac{m_{f}\Phi_{0} - \mu}{kT}\right) \, \, \, .
\label{eq:lambda}
\end{equation}
The equation for the gravitational potential can thus be written as
\begin{equation}
\frac{1}{x}\frac{d^{2}v}{dx^{2}}=
-\frac{3}{2} \beta^{-3/2}\frac{\Lambda}{\Lambda + 1} I_{1/2}\left(\Lambda,\frac{v \beta}{x}\right) \; \;  ,
\label{eq:12}
\end{equation}
with boundary conditions
\begin{equation}
v(0)=\frac{M_{BH}}{M_{\odot}}, \, \, \, \, \, \, v(x_{0})=0 \, \, \, ,
\end{equation}
where $x_{0}=R/a$ is the dimensionless size of the DM halo at which the density and 
pressure of DM vanishes. 
It can be shown that 
for $\Lambda \rightarrow \infty$, equation (\ref{eq:12})
 describes the isothermal gas sphere of King (\cite{king66})
whereas for $\Lambda \rightarrow \infty$,
and $v >> 1$ we recover the Lane-Emden equation for a degenerate FB
(Viollier \cite{viollier94}).
 $I_{n}(k,\eta)$ is a Fermi - like integral defined as
\begin{equation}
I_{n}(\Lambda,\eta)=\int_{0}^{\eta} d\xi \xi^{n}
\frac{1-{\rm e}^{\xi-\eta}}{1+\Lambda {\rm e}^{\xi-\eta}} \, \, .
\label{eq:13}
\end{equation}
The mass density $\rho(r)$ can be written in terms of the 
dimensionless potential $v$ as follows:
\begin{equation}
\rho(r)=\frac{m_{f}^{4}g_{f} 2^{1/2}}{2\pi^{2}\hbar^{3}}
\left(\frac{GM_{\odot}}{a}\right)^{3/2} \beta^{-3/2}
\frac{\Lambda}{\Lambda + 1}I_{1/2}\left(\Lambda,\frac{\beta v}{x}\right).
\label{eq:13d}
\end{equation}
Using $\rho(r)$ and the differential equation for the gravitational
potential $v$, the mass 
enclosed  within a radius $r$ of the DM halo is given
by
\begin{eqnarray}
M(r)&=&\int_{0}^{r}4\pi\rho(r') r'^{2}dr'   \nonumber\\
&=& M_{\odot} \left( -x v'(x) + v(x) - v(0) \right)  \, \, . 
\label{eq:14}
\end{eqnarray}
and the total mass of the DM halo (including $M_{BH}$ and the fermion
halo) is given by
\begin{equation}
M=M(R)=-M_{\odot}v'(x_{o})x_{0} \, \, ,
\end{equation}
where $x_{0}$ is the dimensionless size of the DM halo.
We then study the solutions of equation (\ref{eq:12}) for various values of
$\Lambda$ and $m_{f}$.

\begin{figure}
\vspace{1cm}
\includegraphics[width=90mm]{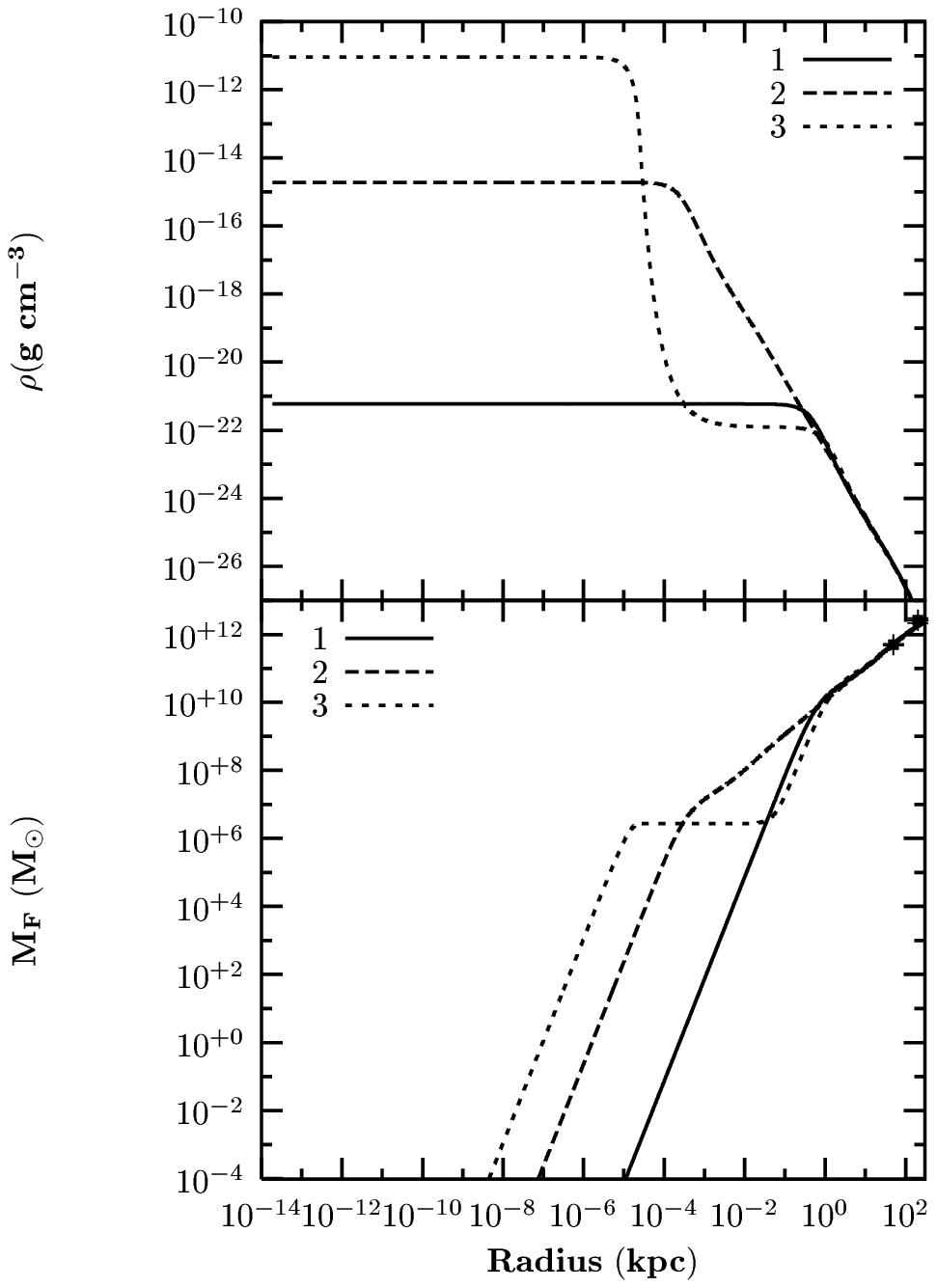}
\caption{The mass density and the mass enclosed  within a radius $r$ are
plotted in the upper and lower panels, respectively.
Near the center of the DM halo, the fermions are completely
degenerate and condensed in a FB of mass $\sim 3 \times
10^{6}M_{\odot}$.
 In the outer edge
of the DM  halo, the density
$\rho$ scales as $ 1/r^{2}$ (van Albada et al.
\cite{albada85}) . The mass of fermions used in this plot is
$m_{f}c^{2} \approx 12 {\rm keV}$ and the size of the FB is about $10^{-2}{\rm pc}$. The data points for the mass
within 50 kpc and 200 kpc are taken from
Wilkinson \& Evans (\cite{evans99}). 
}
\end{figure}

In Fig. 1, we plot the mass density and the mass enclosed 
 for  solutions 
with a $3 \times 10^{12}M_{\odot}$ mass and  
a radius of 200 kpc, which corresponds to our Galaxy.
Throughout our paper, we will study solutions of type 3 which have
degenerate FBs in the center.
For the case of our galaxy,
in order to have a degenerate FB
of $3\times 10^{6}M_{\odot}$
at the center and fit the rotation curve of the Galaxy, i.e.
 reproduce the $1/r^{2}$ law,
the  mass of fermions should be about $m_{f}c^{2} \sim 12 {\rm keV}$.
For a FB of $3\times 10^{3}M_{\odot}$, one would need 
a fermion mass of $m_{f} \sim 450 keV/c^{2}$.
Thus under the assumption of a degenerate FB of
$3\times 10^{3}M_{\odot}$ to $3\times 10^{6}M_{\odot}$ at the center of a DM halo of $3\times 10^{12}M_{\odot}$ with a density scaling as $1/r^{2}$ in the outer edge of the halo, the fermion mass
is constrained in the range
\begin{equation}
 12 \ {\rm keV} \stackrel {\textstyle <}{\sim} m_{f}c^{2}
\stackrel {\textstyle <}{\sim} 450 \ {\rm keV} \, .
\label{eq:limit1}
\end{equation}
Similar solutions for the DM 
distribution in our Galaxy have also been 
studied in Bili\'c et al. (\cite{bmtv02}).

The solutions shown in Fig.~1 have been obtained using the fact that  at large distances the Fermi gas is 
non degenerate. The degree of non-degeneracy is described by equation (\ref{eq:lambda}) which contains the
 chemical potential $\mu$.
Thus, the use of the chemical potential in the  Fermi - Dirac ditribution allows 
to constrain the DM particles in the range 
from 12 to 450 keV, much below the Lee-Weinberg lower limit of about 2 GeV (Lee \& Weinberg \cite{weinberg77}) which was 
established under the assumptions that
DM particles were in thermodynamic equilibrium at the freeze - out temperature.  
We also note that  a suitable candidate for the DM particle as defined by the range 
given in (\ref{eq:limit1})
could be either the gravitino, postulated in supergravity theories with a mass in the  $\sim 1 {\rm keV/c}^{2}$ 
to $\sim 100 {\rm GeV/c}^{2}$
 range (Lyth \cite{lyth99}), or the axino, with a mass in the range between $\sim 10$ and $\sim 100 {\rm keV/c}^{2}$, as predicted by
 supersymmetric extensions of the Peccei-Quinn solution to the strong CP problem (Goto \& Yamaguchi \cite{goto92}).
 For a recent review on DM particle candidates, the reader is referred 
 to Bertone, Hooper \& Silk (\cite{silk04}) and Baltz (\cite{baltz04}).

\subsection{Black hole accretion rate and Pauli principle}

We define the local dynamical (orbital) time scale
and the mass accretion flow as

\begin{equation} 
  t_{dyn}=\frac{r}{v_{ff}} \; \; ,
\end{equation}

\begin{equation}
\dot{M}_{BH}=4\pi r^{2} \rho(r) v_{ff} \;\;\; ,
\label{eq:rate}
\end{equation}
where $v_{ff}$ is the free - fall velocity
\begin{equation}
 v_{ff}=\left(\frac{2Gm(r)}{r}\right)^{1/2} \, , 
\end{equation}
and  $m(r)$ is the mass enclosed within a radius $r$ in the DM halo.

Near the center of the DM  halo, the core becomes 
completely degenerate and the Pauli condition can be written as
\begin{equation}
\left(\frac{g_{f}}{6\pi^{2}}\right)^{1/3}\frac{m_{f}v}{n^{1/3}} = \hbar \, ,
\label{eq:pauli}
\end{equation}
where $n=\rho/m_{f}$ is the number density for DM particles and 
$v$ is the Fermi velocity.
A seed BH of mass $M_{BH}$ that settles at the center of 
the DM halo induces 
a free - fall velocity 
\begin{equation}
v_{ff} = \sqrt{\frac{2GM_{BH}}{r}} \, .
\end{equation}
Replacing $v$ from the equation  (\ref{eq:pauli}) with the last expression 
for $v_{ff}$,  the density of DM at the Pauli limit is given
by 
\begin{equation}
n = \frac{g_{f}}{6\pi^{2}}\frac{m_{f}^{3}}{\hbar^{3}}
\left(\frac{2GM_{BH}}{r}\right)^{3/2} \, \, \, \, .
\end{equation}
We then rewrite  equation (\ref{eq:rate}) for the mass flow 
 using  the Pauli flow condition near the BH
\begin{eqnarray}
\dot{M}_{BH}&=&\frac{8g_{f}}{3\pi\hbar^{3}}m_{f}^{4}G^{2}M_{BH}^{2} \nonumber \\
      &=&1.03\times 10^{-7}g_{f} \left(\frac{m_{f}c^{2}}{15 \ {\rm keV}}\right)^{4}\left(\frac{M_{BH}}{M_{\odot}}\right)^{2} M_{\odot} {\rm yr}^{-1}
\label{eq:199}
\end{eqnarray}
Thus as soon as the region of BH influence is reached, the 
accretion flow becomes  independent of the radius. However, the BH 
accretion rate depends on the fourth power of the fermion mass and on the square of the BH mass.
This dependence is of course  a consequence of the Pauli principle applied 
near the BH.

Here we note that the mass $M_{F}$  of nonrelativistic 
degenerate FBs scales with radius $R_{F}$
as
\begin{equation}
M_{F}R_{F}^{3}=27.836M_{\odot}\left(\frac{15 {\rm keV}}{m_{f}c^{2}}\right)^{8} \left(
\frac{2}{g_{f}}\right)^{2} {\rm pc}^{3} \, .
\end{equation}
The physics of degenerate FBs as a model
of  DM at the center of galaxies was considered in a series of 
papers (Bili\'c et al. \cite{bmtv02},\cite{bmpr02}; Bili\'c, Munyaneza \&
Viollier \cite{bmv99}; Munyaneza, Tsiklauri \& Viollier \cite{mtv98,mtv99};  Munyaneza \& Viollier \cite{mv02};
 Tsiklauri \& Viollier \cite{tv98} and Viollier \cite{viollier94}).
In these papers, it was argued that the supermassive compact objects in
galactic centers are 
FBs rather than BHs.
Our approach in this paper is that   FBs  are not a replacement for
 BHs but are a necessary step for them
to grow    in galactic centers.

In Fig.~2 we plot the mass - radius relation for FBs for 
different values of the fermion mass. In the same plot we have 
drawn the BH line and two horizontal lines that 
correspond to a mass of $3\times 10^{3} M_{\odot}$ 
and $3\times 10^{6}M_{\odot}$.
From this plot, we find an upper limit for the fermion mass of
\begin{equation}
m_{f}c^{2} < 32 \  {\rm MeV} \; .
\label{eq:limit2}
\end{equation}
This upper limit is the value of the fermion mass that would correspond to the
point in the graph where  the BH line crosses the horizontal line
of a mass $M_{F} = 3 \times 10^{3}M_{\odot}$.

\begin{figure}
\includegraphics[width=90mm]{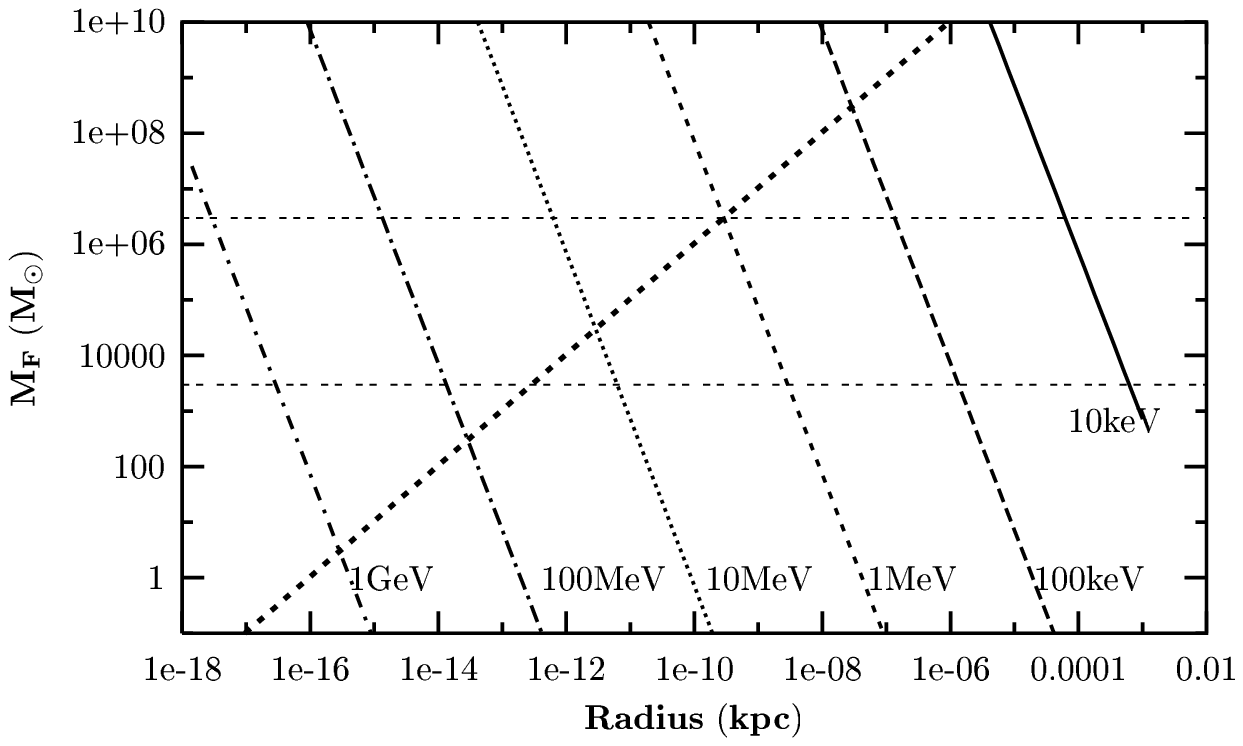}
\caption{The mass - radius relation for degenerate FBs.
The total mass scales as $M_{F} \sim R_{F}^{-3} m_{f}^{-8}$.
In the same plot, we have shown the BH line and two horizontal lines
for the lower and upper limits for the total mass of the degenerate Fermi
balls. Relativistic effects make the mass - radius relation curve  have a 
maximum mass $M_{OV}$ (see equation \ref{eq:ov}) and the size of such a 
FB at the Oppenheimer Volkoff limit is only $4.45 R_{OV}^{s}$ where $R_{OV}^{s}$
is the Schwarzchild radius of the mass $M_{OV}$. The line of BHs and a FB at the OV limit
are nearly indistinguishable in this plot.
}
\end{figure}

The choice of the lowest value for the FB mass
 is motivated by the fact that
once the BH has consumed  a FB of 
 $3\times 10^{3}M_{\odot}$, it  would grow by Eddington - limited
baryonic matter accretion to 
$3\times 10^{9} M_{\odot}$ as observed in 
quasars at redshifts $z \sim 6$.
The choice of  the second value of $3 \times 10^{6}M_{\odot}$, 
is due to increasing evidence for the existence of  a BH of mass $3\times
10^{6}M_{\odot}$ at the 
center of our galaxy. Thus we will assume throughout this 
paper that galaxies which have BHs had
degenerate FBs of masses ranging from $\sim 10^{3}M_{\odot}$
to $\sim 10^{6}M_{\odot}$.
The fermion mass range  given by equation (\ref{eq:limit1})
 is of course
well below  the upper limit for the fermion mass from equation
(\ref{eq:limit2}).

For completeness we note here that stable degenerate FBs exist up 
to a  maximum mass also called the Oppenheimer Volkoff limit
\begin{eqnarray}
M_{OV}&=&0.54195 M_{Pl}^{3}m_{f}^{-2} g_{f}^{-1/2} \nonumber \\
      &=&2.7821 \times 10^{9} M_{\odot} \left(\frac{15{\rm
keV}}{m_{f}c^{2}}\right)^{2}\left(\frac{2}{g_{f}}\right)^{2} \, ,
\label{eq:ov}
\end{eqnarray}
where $M_{Pl}=(\hbar c/G)^{1/2}$ is the 
Planck mass (Bili\'c, Munyaneza \& Viollier \cite{bmv99}).
The size of a such FB at the Oppenheimer Volkoff limit is only
$R_{OV}=4.45R_{OV}^{s}$ where $R_{OV}^{s}$ is the Schwarzschild radius of
the mass $M_{OV}$. Thus there is practically no difference between a black
hole and a FB of the same mass  at the OV limit, as the closest
stable orbit around a non-rotating BH has a radius of $3 R_{s}$ only.

From equation (\ref{eq:ov}), the fermion mass needed to form a degenerate FB of $3\times 10^{9}M_{\odot}$ at the OV limit is about $m_{f} \sim 12 {\rm keV/c^{2}}$.
This value is in agreement with the lower limit on the mass of fermions  obtained from fitting the
Galactic DM halo with a FB of $3\times 10^{6}M_{\odot}$ at the center and $1/r^{2}$ density law from about 3 kpc.
The most compact supermassive object of $3\times 10^{9}M_{\odot}$ has been 
identified as a BH at the center of M87 (Macchetto et al. \cite{macchetto97}).

\begin{figure}
\includegraphics[width=105mm]{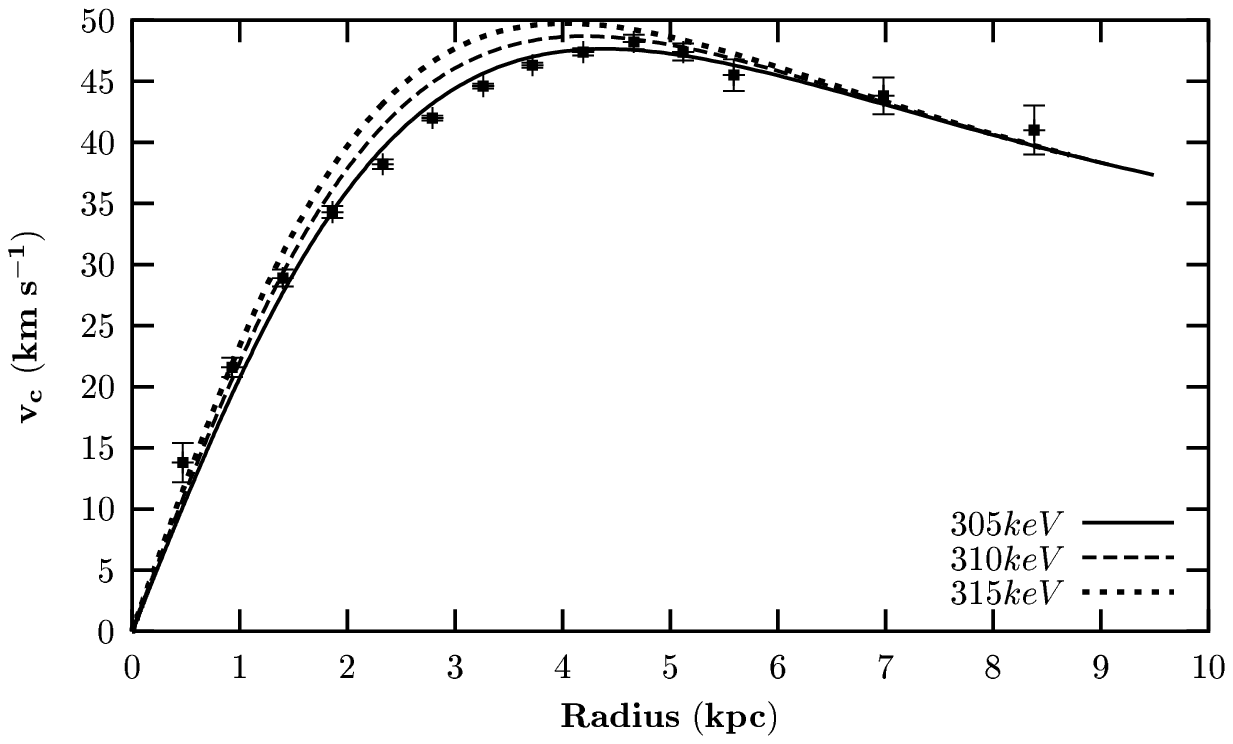}
\caption{Fit of  the rotation curve of dwarf galaxies DD0154. The data points can be 
fitted by a non-degenerate Fermi - Dirac
distribution function for DM particles in the mass 
range discussed in this paper, i.e. for $12{\rm keV} 
\stackrel {\textstyle <}{\sim}m_{f}c^{2} \stackrel {\textstyle <}{\sim} 450 {\rm keV}$. In this plot we have used  a fermion mass
of $m_{f} \sim 305 keV/c^{2}$
to fit the rotation curve of the dwarf galaxy  DD0154. Here we also note that some small
 addition  of baryonic matter
would probably allow a closer  fit to the data.
}
\end{figure}

Apart from fermions as  candidates of DM, bosons have also been considered
as an alternative to the DM constituents in galaxies. Self - gravitating bosons can form boson stars and they are prevented from complete gravitational collapse by the Heisenberg
uncertainty principle $\Delta r \Delta p \sim \hbar$ (Schunck \& Liddle \cite{sliddle77}).
Boson stars are described by Einstein's field equations coupled with the non-linear  Klein Gordon equation
for a complex field $\Psi$ with self - interaction of the form 
$U(|\Psi|)=m_{B}^{2}|\Psi|^{2}+\kappa |\Psi|^{4}$ where $m_{B}$ is the boson mass 
and $\kappa$ is a coupling constant.
It has been shown that boson stars have a maximum mass called the Kaup limit given by
\begin{equation}
M_{Kaup}=0.633 M^{2}_{Pl}/m_{B} \, \, ,
\label{eq:kaup}
\end{equation} 
for $\kappa=0$. Thus for a boson mass of $m_{B} \sim 10^{-23}{\rm GeV/c}^{2}$ the Kaup 
mass limit is about $10^{6}M_{\odot}$ (Torres, Capozziello \& Lambiase \cite{torres00}).
For $m_{B} \sim 1 GeV/c^{2}$ the Kaup limit is as small as $10^{-17}M_{\odot}$.
However if $\kappa \ne 0$, the combination of 
these two free parameters $m_{B}$ and $\kappa$ let boson stars have any mass for any chosen value of the 
boson mass $m_{B}$. It is in this framework that boson stars were attractive as they  could form  very massive objects 
with $m_{B} \ge 1 {\rm GeV/c}^{2}$ ( Schunck \& Mieke \cite{smielke03}).

In Fig.~3, we fit the rotation curve of dwarf galaxies using a Fermi - Dirac
distribution. The rotation curve data points are well fitted in the outer
region. In the inner region, i.e. around 1 kpc, baryonic matter is perhaps
needed to get a better fit of the rotation curve. In order to fit the rotation curve of 
dwarf galaxies, a non-degenerate FD distribution is needed. 
The data points are taken from  Carignan \& Purton (\cite{cp98}) and the mass
of the the dwarf galaxy DD0154 is taken to be $3\times 10^{9}M_{\odot}$ with
a size of $\sim 10 \ {\rm kpc}$ (Kravtsov et al. \cite{primack98}).
The merging of dwarf galaxies would generate another galaxy with a degenerate
FB, and the BH would grow by the same mechanism considered in this
paper in the merged galaxy.
The 
 Fermi - Dirac distribution studied in this paper allows  two types of 
 solutions: One which fits dwarf galaxies with a non - degenerate core
  and the other solution with $1/r^{2}$ drop off in the outer edge 
 of the halo.
 Degenerate FBs are  consistent with data on the rotational 
curves only in the second type of solution with  a $1/r^{2}$ density profile at large
 distances.

\section{Black hole growth}

\subsection{Inner dark matter accretion and the quantum cascade mechanism}
In this section we will consider the accretion of
inner DM, i.e.
DM  inside the FB and in the next section we will analyse the accretion of  DM outside the FB.
In the last section, we established  that the inner part of the the Fermi ball is completely degenerate and 
have used  the  Pauli
Principle to obtain  the BH consumption rate given by equation (\ref{eq:199}), which can  be integrated to give the BH mass as a function of time:
\begin{equation}
	M_{BH}=M_{\odot}\left[\frac{M_{\odot}}{M_{BH}^{i}}-\frac{1}{t_{0}}\left(t-t^{i}\right)\right]^{-1},
\label{eq:111}
\end{equation}
where $M_{BH}^{i}$ is the mass of the seed BH at the starting time $t^{i}$
of accretion, and $t_{0}$ is a time parameter given by
\begin{equation}
t_{0}=9.7\times  10^{6}{\rm yr} \left (\frac{15 {\rm keV}}{m_{f}c^{2}}\right)^{4}
g_{f}^{-1}
\; .
\label{eq:112}
\end{equation}
The last two equations define the growth of a seed BH immersed at the center
of a DM halo with a degenerate FB from which it is seen that
 the BH mass strongly  depends
 on  the fermion mass.

  From  equation (\ref{eq:111}), we can estimate the time 
 $t_{\infty}$ needed for the BH mass to be infinite
 to be
 \begin{equation}
  t_{\infty}=t^{i} + \frac{M_{\odot}}{M_{BH}}t_{0} \; ,
 \end{equation}
 which for an initial seed of $5 M_{\odot}$ at $t^{i}=0$ would give
 \begin{equation}
 t_{\infty} = 1.94 \times 10^{6} {\rm yr} \left (\frac{15 {\rm keV}}{m_{f}c^{2}}\right)^{4}g_{f}^{-1} \; .
\end{equation}
For a fermion mass of $m_{f}=1 {\rm keV/c}^{2}$ the above timescale $t_{\infty}$ would be 
about $5 \times 10^{10}$ years whereas
 a fermion mass of $m_{f} = 1 {\rm MeV/c}^{2}$ would give a growth timescale of  about $t_{\infty}=5\times 10^{-2}$ years.
 However, the BH mass cannot be greater than the mass of the FB that feeds it and we will require that the BH mass
 will grow from DM accretion only up to the mass of the FB and this will constitute 
 one of the main assumptions in this paper.  Once the BH mass has grown to the FB ball mass, a further growth to
 $10^{6-9}M_{\odot}$ is  achieved via Eddington - limited baryonic matter accretion. 

 The mechanism of consumption of DM particles by the BH can be understood
using a quantum cascade mechanism. First, low angular momentum
DM particles at the inner Fermi surface in phase space will be consumed by the
BH. Due to a high degeneracy pressure, high angular momentum particles will
be pushed to the inner Fermi surface and taken up by the BH. Outside the 
FB, fermions
will hit the outer Fermi surface and  for the FB to be hit continuously by DM particles, we
use  molecular clouds as perturbers of DM orbits. Thus, we have a three body interaction between the FB, the 
molecular clouds and DM particles.
For our mechanism of consumption of DM by the BH to work,  we will assume that the dark matter particles interact with
the FB through inelastic collisions.
In hitting the FB, some of the DM particles will be excited  to  a slightly higher energy level
and get 
stuck in high
momentum segments of phase space.
This process will continue until the FB has been 
  completely consumed   by the
BH and thereafter baryonic matter accretion will control the BH growth.

\subsection{Outer dark matter accretion}
We study  the growth of a seed BH of 
about $5 M_{\odot}$ introduced at the center of 
DM halos of masses ranging from $10^{9}M_{\odot}$ to $10^{14.5}M_{\odot}$. The lower limit corresponds to dwarf galaxy halo masses while the upper limit corresponds to the most  
massive halo such as that of M87.
The gravitational potential and the mass
density of DM
are given by Eqs. (\ref{eq:12}) and (\ref{eq:13d}), respectively. 
It is well established that the BH will induce a density  law
$\rho(r) \sim  r^{-3/2}$ at  the center of a DM 
halo (Lightman \& Shapiro \cite{lightman78}; Quinlan et 
al.  \cite{quinlan95}; MacMillan \& Henriksen \cite{macmellan02}; Merritt
\cite{merritt03}).
Following Spitzer \& Schwarzschild (\cite{spitzer51,spitzer53}) 
and Julian (\cite{julian67}), we assume  that 
DM orbits outside the degenerate FB
 are perturbed by molecular clouds  of 
mass $m_{pert} \approx 10^{6}M_{\odot}$.
 It  has been recently  reported that
molecular gas of total mass $\sim 10^{10}M_{\odot}$ is 
observed in the host galaxy of the same  quasar SDSS J114816.64+525150.3 at 
 $z \approx 6.41$ (Walter et al. \cite{walter03}; Carilli et al. \cite{carilli04}).

In this paper, we assume that there is a compact degenerate FB at
the center of galaxies that have BHs.
The seed BH first consumes  the degenerate FB and grows to a mass of about $10^{3\, - \, 6} M_{\odot}$ and then grows by
 Eddington - limited baryonic
matter accretion to higher masses of $10^{6 \, - \, 9}M_{\odot}$.


We will be mostly concerned with the calculation of the time needed to refill the BH loss cone using
molecular clouds as DM perturbers. To this effect, we 
 denote by
 $r$  the radial distance from the  BH to the DM particle position,
$ \tilde{r}$  the distance from the BH to the molecular clouds,
 $\theta$ is the angle between the direction from the BH to  the DM particle and 
its projection in the disk and   $\varphi$ is the angle 
between the direction from the BH  to  the molecular clouds and to  
the foot of the DM particle location in the disk. The diagram for the
geometry used is shown in Fig.~4.

In order to get the expression for the 
 time $\tau_{refill}(r)$ needed to refill the loss 
cone 
which is  due to multiple small-angle 
gravitational (Coulomb) encounters, 
 we evaluate the 
random walk integral (Spitzer \cite{spitzer87}) 
\begin{equation}
\Delta v^{2}=\int_{\tilde{r}_{min}}^{\tilde{r}_{max}}\int_{0}^{\pi}
\left(\frac{2vp_{0}}{p}\right)^{2} \, \tilde{r} \,  d\tilde{r} \, d\varphi
\, n_{pert}(\tilde{r}) \, v \, \tau_{refill}(r) \, \,\, ,
\label{eq:001}
\end{equation}
where $v$ is the relative velocity of DM particle and molecular clouds at
large separation, $p_{0}$ and $p$ are given by
\begin{equation}
p_{0}=\frac{G(m_{pert}+m_{DM})}{v^{2}}, \, \, \, \, \, \, \, \, \, \, \,
p^{2}=r^{2}+\tilde{r}^{2}-2r\tilde{r} \cos \theta \cos \varphi\,\, .
\end{equation}

\begin{figure}
\includegraphics[width=90mm]{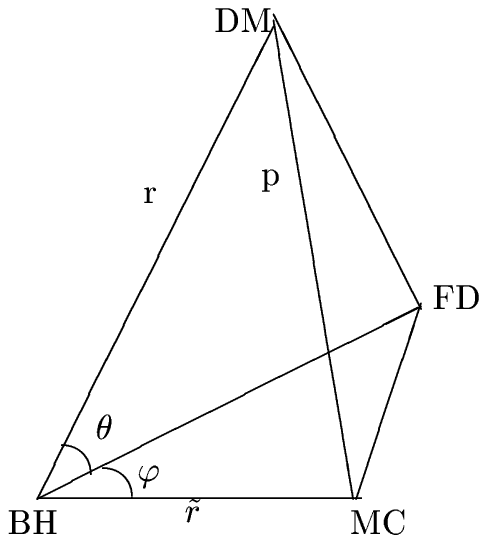}
\caption{Geometry for the system BH (BH), DM (DM) particles and
molecular clouds (MC). FD stands for the foot (projection) of the  DM
particle in the disk. The angles BH $\; - \;$ FD $\; -\; $ DM and
 MC $\; - \;$ FD $\; -\; $ DM  are both right angles. 
}
\end{figure}

In the above equations, we have  used the values of 0.1 kpc and 20 kpc 
for $\tilde{r}_{min}$ and $\tilde{r}_{max}$, respectively.
We will also  assume the following 3D number density distribution equivalent
to the surface density for molecular
clouds (G\"usten \& Mezger  \cite{gusten82})
\begin{equation}
n_{pert}(\tilde{r})=\frac{\sigma_{0}}{2\tilde{r}}\exp\left(-\frac{\tilde{r}}{r_{0}}\right)\,\,
 ,
\label{eq:002}
\end{equation}
where $r_{0}=5 \ {\rm kpc}$ is a scaling distance and $\sigma_{0}$ is a constant
denoting  the surface
density to be chosen a posteriori.
Inserting the last expression (\ref{eq:002}) for the number density into
equation (\ref{eq:001}), after integration we get the following equation for
the time to refill the loss cone:
\begin{equation}
 \tau_{refill}(r)= \frac{v^{3}r_{0}}{2\pi G^{2}\left(m_{DM}+m_{pert}\right)^{2}
\sigma_{0}I(r,\theta)} \left(\frac{\Delta v}{v}\right)^{2} \, \, ,
\label{eq:003}
\end{equation}
with
\begin{equation}
 \left(\frac{\Delta v}{v}\right)^{2} =\left(\frac{R_{s}}{r}\right)^{2}=
\left( \frac{2GM_{BH}}{c^{2}r}\right)^{2} \,\, ,
\end{equation}
where the function 
$I(r, \theta)$ is given by  the following integral
\begin{equation}
I(r,\theta)=\int_{x_{min}}^{x_{max}}dx
\frac{\exp(-x)}{\sqrt{x^{4}-2(\frac{r}{r_{0}})^{2}\cos(2\theta)x^{2}
+(\frac{r}{r_{0}})^{4}}} \,\, ,
\end{equation}
In the last integral, $x_{min}=\tilde{r}_{min}/r_{0}=0.02$ and
$x_{max} = \tilde{r}_{max}/r_{0}=4$. The function $I(r,\theta)$ tends to a
constant for $r\rightarrow 0$ whereas for large values of $r$ it scales as
$1/r^{2}$.

\begin{figure}
\includegraphics[width=90mm]{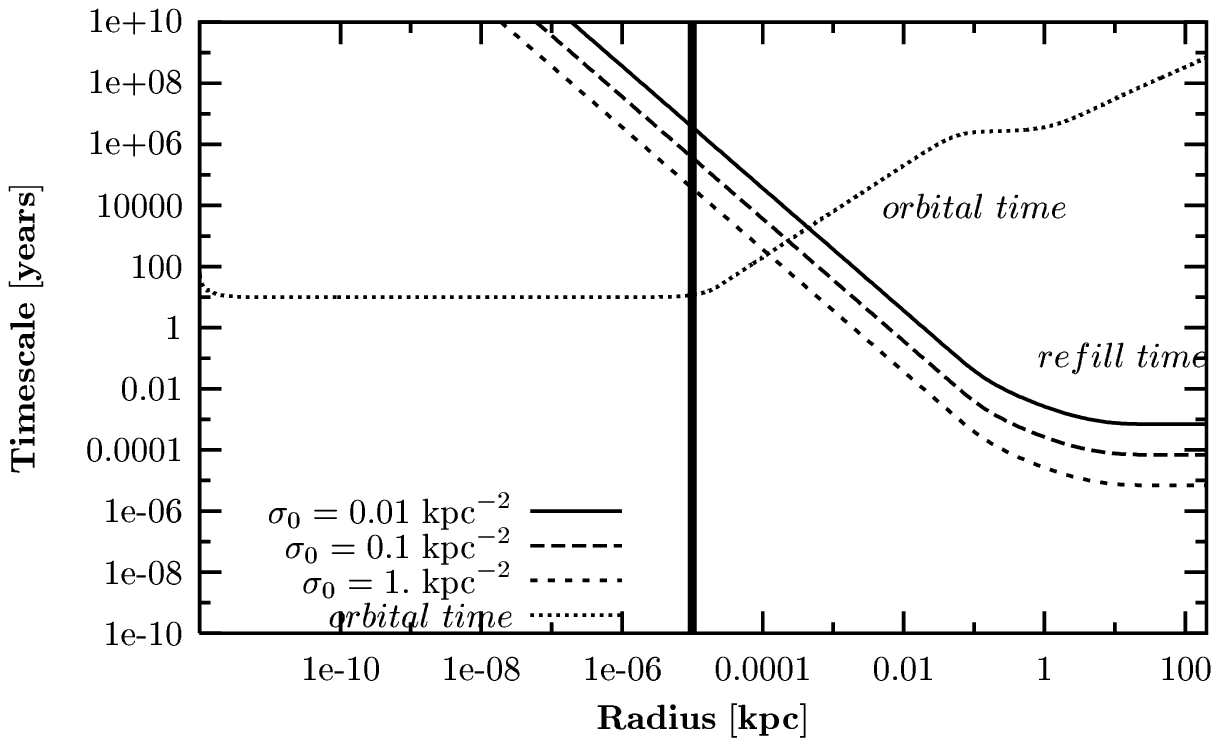}
\caption{The refilling and orbital timescales plotted as a function of the radius from 
the center of the DM halo. In this plot,
 the mass of the FB
is $3\times 10^{6}M_{\odot}$ and the corresponding fermion mass is $12{\rm keV/c^{2}}$.
For small $r$, the time to refill the loss cone is much greater than the age of
the Universe. For the growth of the BH, we use the orbital time scale and the BH feeds 
on DM from inside the FB.
 For the DM inside the FB to be fully consumed 
by the BH, we introduce a quantum cascade effect which is a result
of the existence of a high degeneracy pressure that pushes high angular
momentum particles to the inner Fermi surface.
The particles at the inner Fermi surface in 6D phase space will then be consumed by the
BH. This process continues until the BH has consumed the entire
FB. The BH then grows by Eddington -limited baryonic matter accretion. 
The FB is fed from DM from outer orbits but its accretion rate 
is much lower than that of the BH inside the FB as seen in the next figure. The vertical thick line shows the size of the FB.
}
\end{figure}

\begin{figure}
\includegraphics[width=90mm]{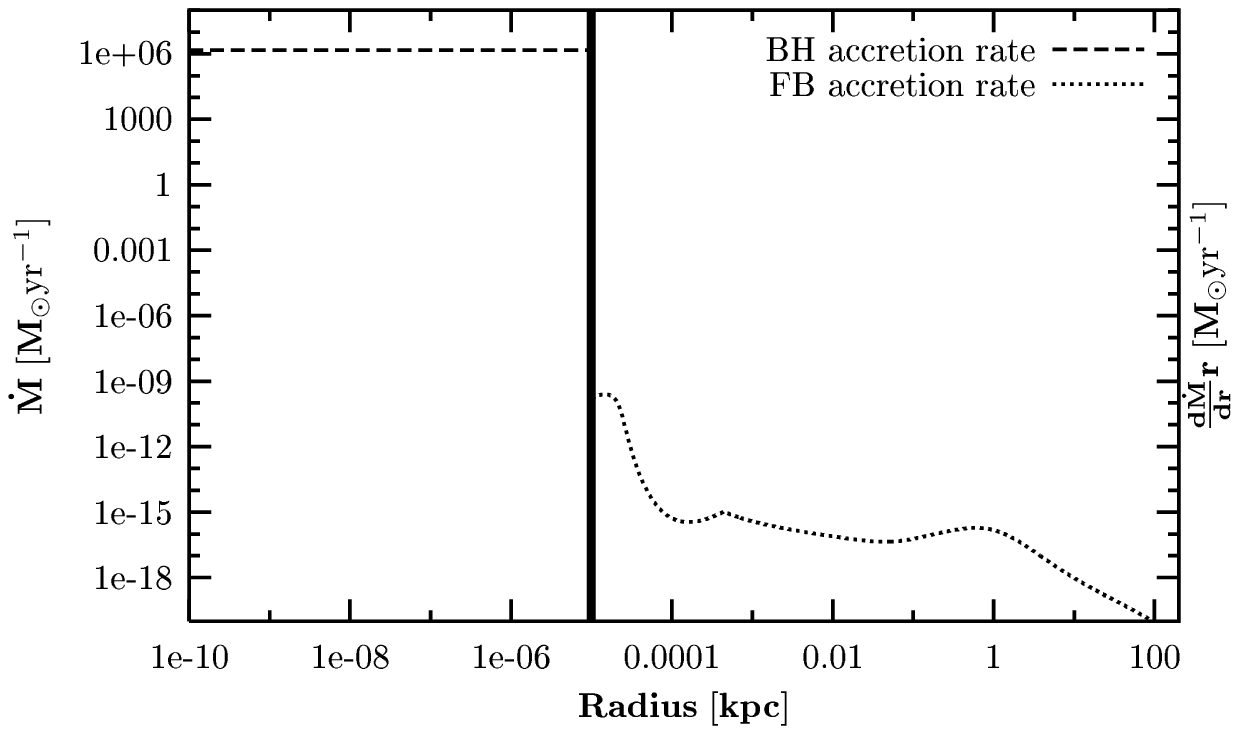}
\caption{
The mass accretion flow as a function of distance from the center.
For a given mass of the BH, the accretion rate is independent of the
radius until the Bondi radius where the BH dominates the gravitational
potential of the fermions. As the BH mass approaches that of the  Fermion 
ball, the BH accretion rate becomes constant until the radius of the FB and  as the size of the FB is reached, 
the DM accretion rate onto the FB drops dramatically. Thus the growth rate of the 
FB is much lower than that of  the BH which grows by feeding on DM from inside the FB. 
Similarly to the last plot, we have used here a FB mass of $ 3\times 10^{6}M_{\odot}$.}
The left side of the plot shows the BH accretion rate and  the FB consumption rate is shown on the right side of the figure.

\end{figure}

\begin{figure}

\includegraphics[width=90mm]{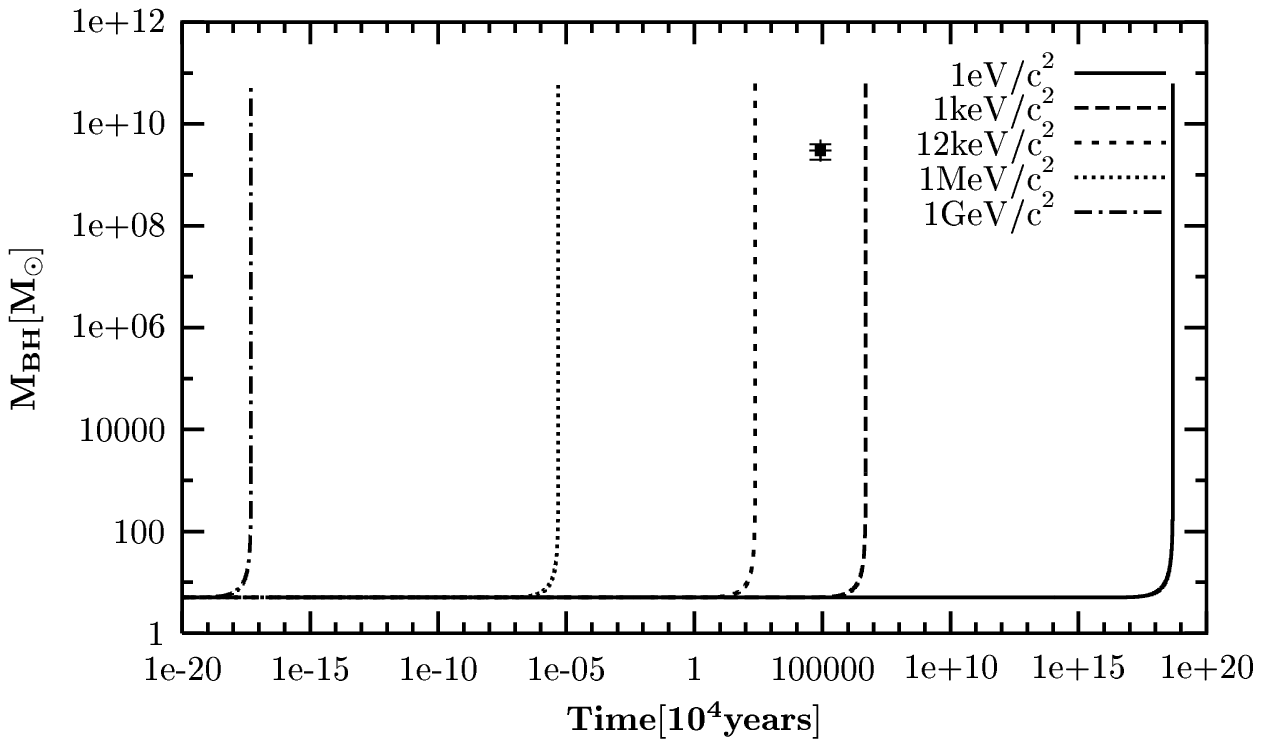}
\caption{The black hole mass growth from only DM accretion. The fermion mass is varied as shown on the plot.
It can be seen that the BH mass can  grow to $10^{3-4}M_{\odot}$
in about $10^{7-8}$ years if the fermion mass is in the range between  about 1 keV and 1 MeV.
For fermion masses $m_{f}   \stackrel {\textstyle <}{\sim} 1 {\rm keV/c}^{2}$, the formation of a BH of $10^{3}M_{\odot}$ would require a timescale much greater than
than  a Hubble time. On the other hand, for fermion masses 
$m_{f}  \stackrel {\textstyle >}{\sim} 1 {\rm MeV/c}^{2}$, 
a $10^{3}M_{\odot}$ BH would have formed in less than
than one  year after the beginning of the Universe. Thus the degeneracy pressure of fermions allows for the capture of 
a large amount of mass if the fermion mass is constrained in 
a range between  about 1 keV and 1 MeV. The data point shown at $t=8.47\times 10^{8}$ years  is the recent detection of a BH mass
of $3\times 10^{9}M_{\odot}$ BH in the quasar SDSS SDSSJ114816.64+525150.3.}
\end{figure}

The FB is consumed  from inside by the BH with an accretion rate 
\begin{equation}
\dot{M}_{BH} = 4 \pi r^{2} \rho v_{ff} , \; \; \; \; r < R_{F} , 
\end{equation}
which after using the Pauli degeneracy condition is given by equation
(\ref{eq:199}). Thus  the BH gains by consuming DM inside the FB and of course the FB loses its DM particles.
On the other hand the FB is also fed on the outside by 
direct collision via the loss cone mechanism.
The FB growth rate  is given by
\begin{equation}
\dot{M}_{FB} =
\int_{R_{F}}^{R_{DM}} 4 \pi r^{2} \rho(r) \delta \Omega(r) \tau(r)^{-1} dr \; \; \; \; \; r > R_{F}, 
\end{equation}
where $R_{F}$ is the size of the FB, $R_{DM}$ is the DM halo size
 and $\delta \Omega$
is the fraction of the solid angle subtended  by the loss cone 
and is given by

\begin{equation}
\delta \Omega= \pi \left(\frac{R_{s}}{r}\right)^{2}=\pi
\left( \frac{2GM_{BH}}{c^{2}r}\right)^{2} 
\end{equation}

The timescale $\tau$ for the FB growth  is given by
\begin{eqnarray}
\tau =
\left\{
\begin{array}{l}
 \displaystyle{t_{dyn}  \hspace{3cm} \; , \; \; \ t_{refill} < t_{dyn} } \; \;  \\ [.5cm]
 \displaystyle{ t_{refill}\hspace{3cm} , \; \; t_{dyn} < t_{refill} } \; \; , \\
\end{array} \right.
\label{eq:time}
\end{eqnarray}
where $t_{dyn}$ and $t_{refill}$ are the dynamical (orbital) and loss cone  refilling timescales, respectively.
As soon as the FB is consumed, the capture radius for DM particles coming from the 
outside drops rapidly.

In Fig.~5, we plot the dynamical and the refilling timescales as functions
of the radius. At small radii, i.e. $r < R_{s}$, the time
$t_{refill}$ to
refill the loss cone is too long for the DM to feed the BH.
Near the BH, we use the dynamical timescale to feed the BH.
For radii greater than the size of the FB, we will use the slowest of the  two timescales
to feed the FB. Thus the FB feeds from the outside  DM particles via the loss cone mechanism while the BH feeds 
from DM inside the FB via the quantum cascade mechanism.


In Fig.6, we plot the mass flow as a function of the radius from the center.
To this end, we fix the mass  of the FB to $3\times 10^{6}M_{\odot}$ and
 use a fermion mass of $m_{f} \sim 12 {\rm keV/c}^{2}$.
Using the BH accretion rate equation (\ref{eq:199}), the
accretion rate becomes $\dot{M} \sim  10^{6} M_{\odot}{\rm yr}^{-1}$
 for a fermion mass of $m_{f}=12 {\rm keV/c}^{2}$ and a 
BH mass of $3\times 10^{6}M_{\odot}$.
For fermions of mass $450 \  {\rm keV/c}^{2}$ and 
a BH mass of $3\times
10^{3}M_{\odot}$, the accretion rate is 
$\dot{M} \sim  10^{4}M_{\odot}
{\rm yr^{-1}}$.
It is seen than the BH accretion rate is independent on the radius and drops 
significantly when the size of the FB is reached. Thus the BH grows much faster from 
the inner  DM particles than the FB from the outer DM particles.
After the BH has reached a mass equal to that of the FB, i.e. $10^{3-4} M_{\odot}$, 
its growth is 
controlled by Eddington baryonic matter accretion.


In Fig.~7, we plot the BH mass as a function of time for different values of 
the fermion mass. The BH mass - time dependence is
given by equation (\ref{eq:111}). It is shown that for fermion masses lower than about 1 {\rm keV}, it would take more 
than $10^{10}$
years for the  BH to grow to $10^{3-4}M_{\odot}$. However, due to the Pauli principle used in
the  derivation of equation (\ref{eq:111}), 
if the fermion mass is constrained in the mass range between about 1 {\rm keV}
and 1 {\rm MeV},  a seed BH of 5 $M_{\odot}$ can  
grow to a $10^{3-4}M_{\odot}$ BH in about $10^{7-8}$ years, which would then 
grow by Eddington - limited baryonic 
accretion to $10^{9}M_{\odot}$ at  a redshift of $z \approx 6.41$ (see next sub - section).
Fermion masses greater than about 1 MeV would make the BH grow much faster and the BH could be as large as $10^{3-4}M_{\odot}$ in less than
about one year after the beginning of the Universe.

  In the next section, we will  discuss
the Eddington baryonic matter accretion and discuss
 the plots of  BH growth  from both
DM and Eddington baryonic matter accretion.

\subsection{Eddington limited baryonic accretion}

\begin{figure}
\includegraphics[width=90mm]{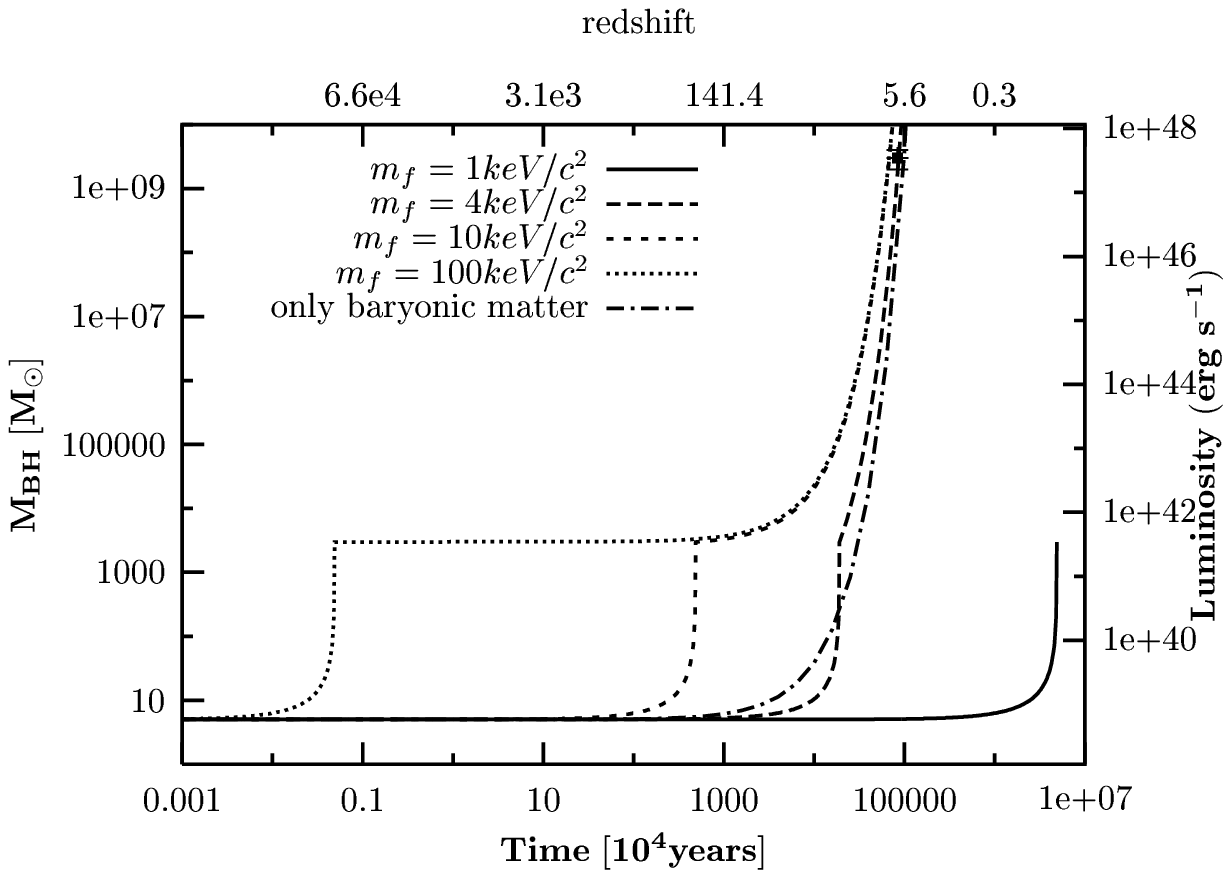}
\caption{The mass  $M_{BH}$ of the BH is plotted as a function of 
time $t$ for DM  and Eddington - limited baryonic matter accretion. 
It is seen from this plot that DM dominates the BH growth at early times. 
The mass of fermions is varied as shown on the plot.
The total mass of the degenerate  FB has a mass of
$3\times 10^{3}M_{\odot}$.
We find that  fermion masses of  $m_{f} 
\stackrel {\textstyle >}{\sim}
 4{\rm keV/c}^{2}$ suffice for 
stellar seed BHs to grow to about $10^{3}M_{\odot}$ which then grow by Eddington - limited
baryonic matter accretion to about $3\times 10^{9}M_{\odot}$. 
The obtained constraint on the fermion mass covers the range of dark matter masses
discussed earlier. The data point shown at $z=6.41$ corresponds to the BH mass of $\sim 3\times 10^{9}M_{\odot}$
in  the quasar SDSSJ114816.64+525150.3.
}
\end{figure}

\begin{figure}
\includegraphics[width=90mm]{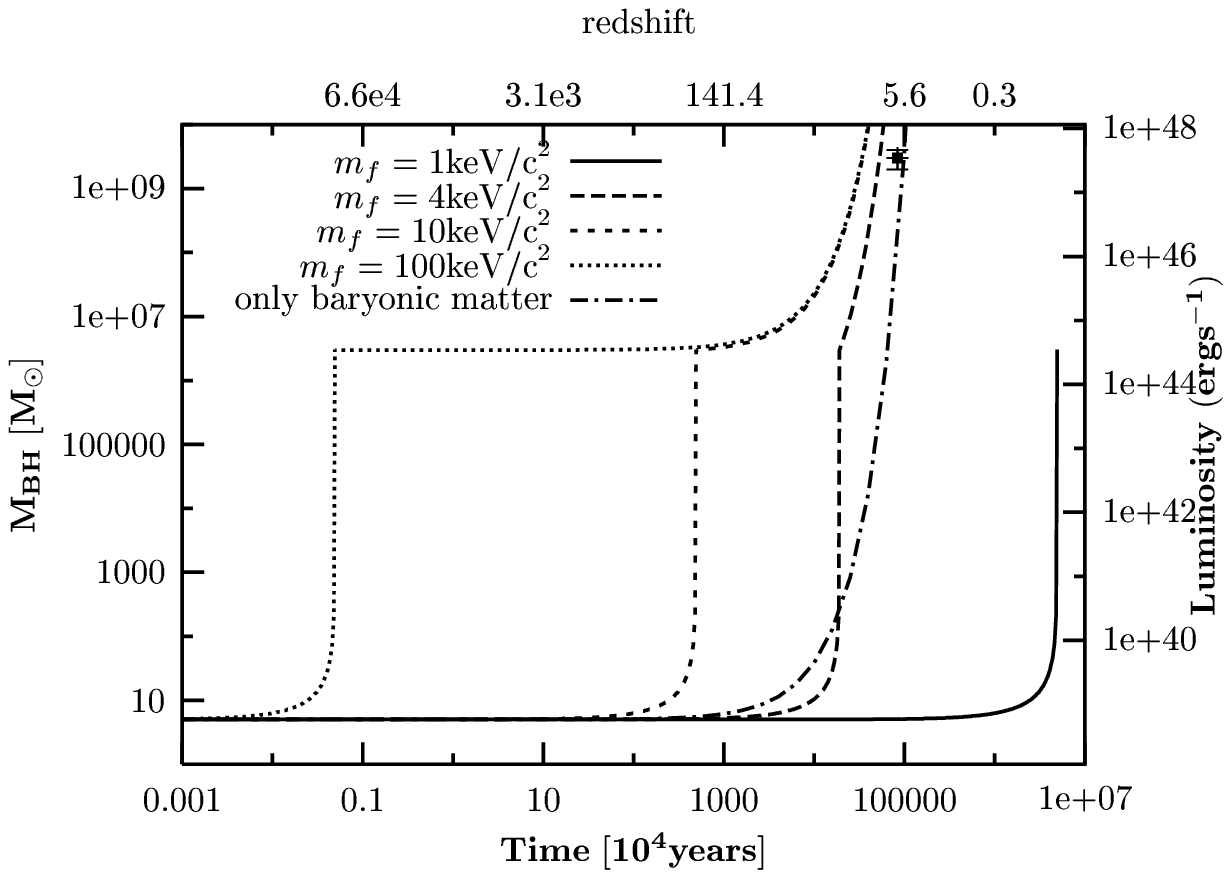}
\caption{The same as in the last figure but for a degenerate FB mass of
$3\times 10^{6}M_{\odot}$.}
\end{figure}

The details of Eddington - limited baryonic  accretion have been studied by many authors,
 here we 
refer the reader to  Wang \& Biermann (1998) and  provide just the main formulae:
\begin{equation}
\frac{dM_{BH}}{dt}=\frac{M_{BH}}{\tau_{EDD}} \, ,
\label{eq:38}
\end{equation}
where $\tau_{EDD}=10^{7.69} {\rm years}\times \left(\frac{\epsilon}{0.1}\right)$ is the corresponding time scale, $\epsilon$ being the efficiency. This value for $\tau_{EDD}$ has been chosen to fit the exponential growth of BHs 
given by Fig.~1  of Wang \& Biermann (\cite{wang98}).
The BH mass is found to grow in this case as
\begin{equation}
M_{BH}(t) = M_{BH}^{i}{\rm exp}\left(\frac{t-t^{i}}{\tau_{EDD}}\right)\, ,
\label{eq:39}
\end{equation}
and the corresponding Eddington luminosity for an efficiency $\epsilon$ of 0.1 is
\begin{equation}
L = \epsilon {\dot M_{BH}}c^{2}  
  = 0.1 \frac{M_{BH}}{\tau_{EDD}}c^{2} \, .
\label{eq:40}
\end{equation}

In Fig.~8 and 9, we present the growth of supermassive BHs accreting DM and baryonic
matter. It is seen from the plot that 
a BH grows faster in the beginning  accreting  DM rather than baryonic matter.
In Fig.8, the degenerate FB has a mass of  $3\times 10^{3}M_{\odot}$
whereas we use a  FB mass of $3\times 10^{6}M_{\odot}$ in Fig.~9.
The data point shown in Fig.~8 \& 9 at $t=8.4\times 10^{8} \ {\rm years}$ is the 
recent detection of  a supermassive BH of mass of 
$M_{BH} \sim 3\times 10^{9}M_{\odot}$ in the 
quasar SDSS J114816.64+5251 at a redshift of $z=6.41$ (Willott, McLure \& Jarvis \cite{willot03}; Fan et al. \cite{fan01}).

 It is seen from Fig.8 \& 9 that
only baryonic matter accretion
 cannot  fit this data point and the 
curve for the growth of the BH due to baryonic 
matter  lies on the right side of the data point. 
Using 
equation (\ref{eq:39}) the total mass
of the BH from Eddington - limited baryonic accretion at $t=8.4\times10^{8}$ years
i.e. $z=6.41$ is $1.4\times 10^{8}M_{\odot}$ which is about ten times less
than the inferred BH mass of $3\times 10^{9}M_{\odot}$, clearly outside the error
range.
The growth of a seed BH from 5 $M_{\odot}$
to $3\times 10^{9}M_{\odot}$ 
 is accomplished in two steps:
First, the BH  completely consumes    the degenerate Fermi  core with 
 masses  between $3\times 10^{3}M_{\odot}$ and $3\times
10^{6}M_{\odot}$.
In the second step, the $10^{3}-10^{6}M_{\odot}$ BH feeds on 
baryonic matter  accretion  to reach the mass
of  $3\times 10^{9}M_{\odot}$ at a redshift of $z \sim
6.41$.

\begin{figure}
\includegraphics[width=90mm]{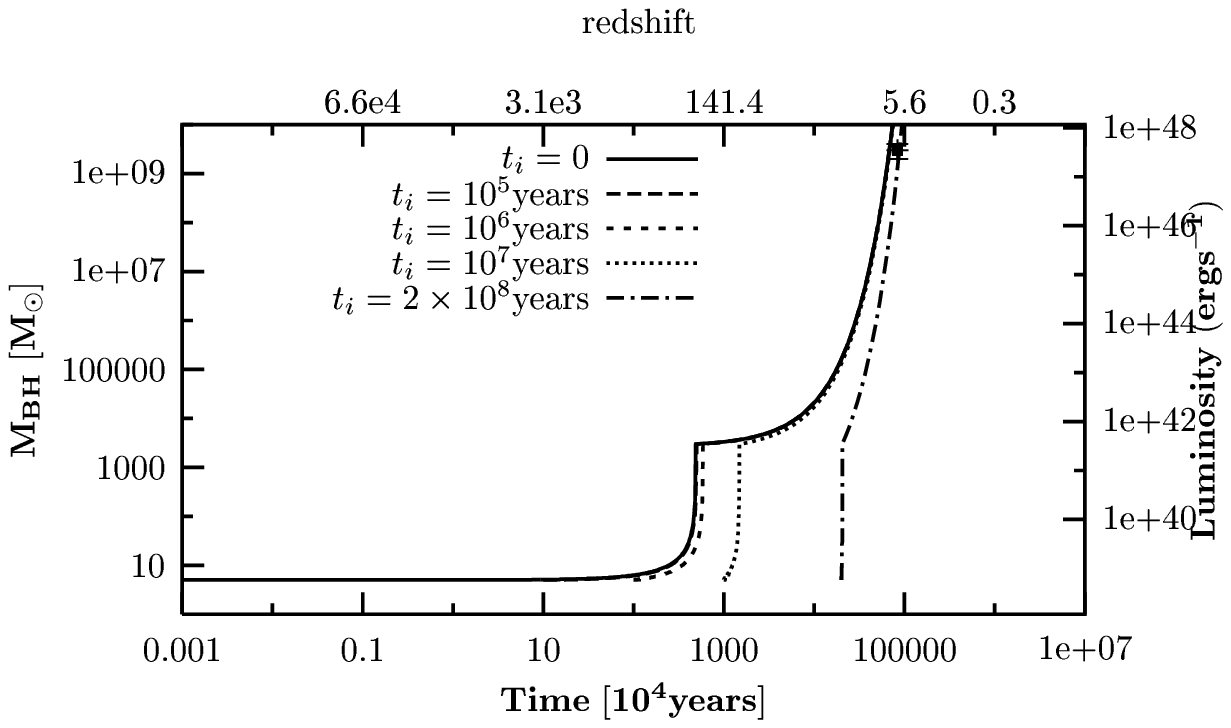}
\caption{The BH growth dependence on the 
starting time of accretion. We find that starting DM accretion at 
a time of $2\times 10^{8} {\rm years}$ would fit the recent 
discovery of a  $3\times 10^{9}M_{\odot}$ BH in the quasar SDSSJ1148+5251 at 
a redshift of $z=6.41$, which is the most distant known quasar, observed only 840 million years after the beginning of 
the Universe (Willot, McLure \& Jarvis \cite{willot03}).
The fermion mass used in this plot is $m_{f} = 10 {\rm keV/c}^{2}$. However, any higher
fermion mass would more sharply and rapidly raise the mass of a seed BH from 5
$M_{\odot}$ to $\sim 10^{3}M_{\odot}$. 
}
\end{figure}

In Fig.~10, we investigate how the 
growth of the mass of the BH depends on the  accretion starting time. 
To this effect, we fix the total mass of the FB to
$3\times 10^{3}M_{\odot}$ and  the fermion mass to $m_{f}= 10 {\rm keV/c}^{2}$ and vary
the initial time when the DM accretion starts as shown on the
graph.
We find that the data point of $3\times 10^{9}M_{\odot}$ 
at $z \sim 6.41$ is well fitted for an accretion starting time of
$t \sim 2 \times 10^{8} {\rm years}$ for any fermion greater
than   $m_{f}  \stackrel {\textstyle >}{\sim} 10
{\rm keV/c}^{2}$.
Here we also note that Eddington baryonic matter accretion
could also allow the seed BH   
to grow to a  $3\times 10^{9}M_{\odot}$ BH if the efficiency is 
$\epsilon=1.1\times 10^{-2}$. The accretion  starting time
of $t \sim 2 \times 10^{8} {\rm years}$ 
corresponds to a redshift 
of reionization of $z\approx 17$ as obtained by WMAP 
observations (Krauss \cite{krauss03}).


\begin{figure}
\includegraphics[width=90mm]{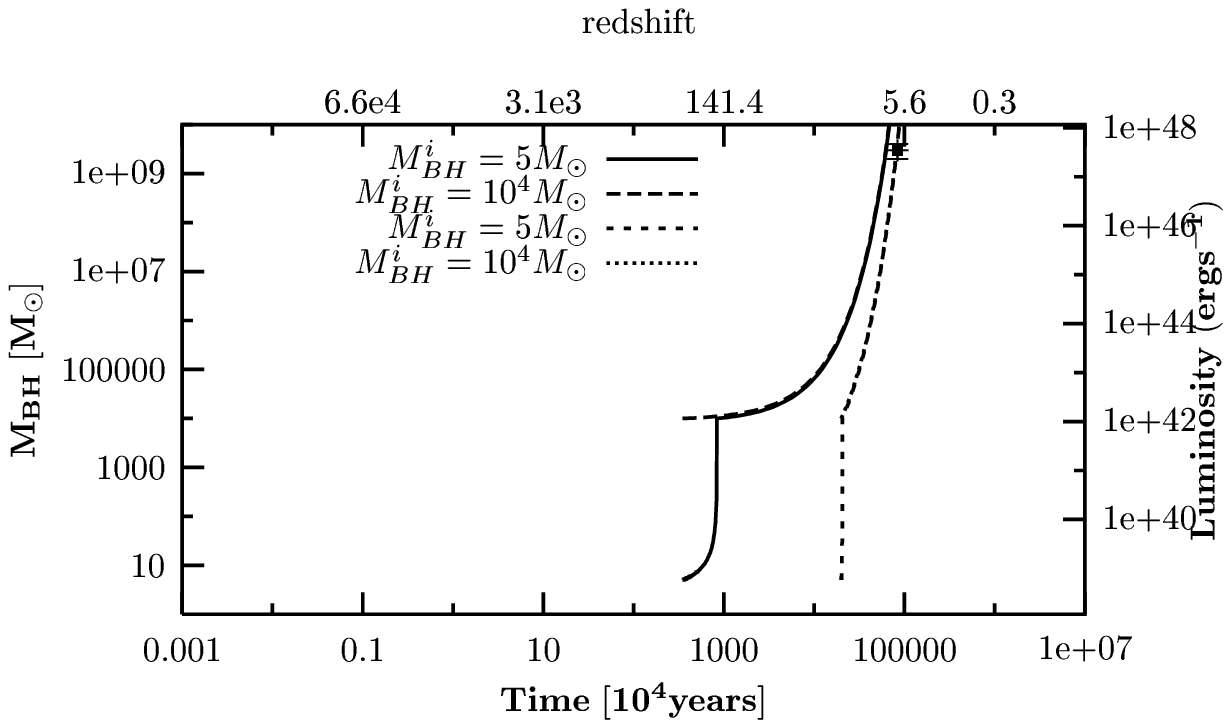}
\caption{ We investigate how the BH growth depends on the mass of the seed BH.
To fit the data point of mass $M \sim 3 \times
10^{9}M_{\odot}$ at $z \approx 6.41$,
the seed BH cannot be greater than $10^{4}M_{\odot}$ for a FB
of $10^{4}M_{\odot}$ and the accretion starting time should be about
$z \approx 17$.
In this plot, the fermion mass is fixed to $m_{f} \approx 10 {\rm keV/c}^{2}$.
Thus, any other mechanism that could create massive seed BHs would generate the same growth.
}
\end{figure}

Finally in Fig.~11, we investigate how the growth of the BH is 
affected by the mass of the seed BH  at different 
starting times of accretion.
In order to constrain the mass of the seed BH, we vary the mass of the
FB from $10^{3}M_{\odot}$ to $10^{6}M_{\odot}$ and also
  the mass of the seed BH from $5 M_{\odot}$ to $10^{4}M_{\odot}$
at initial times of $t^{i}=3.5\times 10^{6}$ and $2 \times 10^{8}$ years.
These values
correspond to the earliest and latest plausible times
for the formation of the first stellar seed BHs in order to grow into
supermassive BHs at redshift $z \approx 6.41$.
  From recent WMAP observations (Spergel et al. \cite{spergel03}), BHs of
masses $\sim 10^{5}M_{\odot}$ might have formed at a redshift $z \geq 20$. 
In Fig.~11, we have 
plotted only the limiting curves 
which show that 
the mass of the FB 
should be $M_{F} \le 10^{4}M_{\odot}$ and the mass of
the seed BH cannot be greater than about $10^{4}M_{\odot}$.
The best fit is obtained for a BH formation time  a reionization time
of $t^{i} \sim 2\times 10^{8} {\rm years}$, i.e. $z \approx 17$.


 It is interesting to estimate the maximum  
luminosity attainable by an accreting BH of this mass. This is usually 
considered to be the Eddington luminosity at 
which the outward radiation pressure equals the inward 
gravitational attraction. 
Using equation (\ref{eq:40}),  the luminosity of a 
$3\times 10^{9}M_{\odot}$ BH is of the order of a 
few $10^{47} {\rm erg \  s}^{-1}$ and the corresponding accretion rate is 
about $65 \ M_{\odot}/{\rm yr}$. 
However, such high luminosity values (Fan et al. \cite{fan01}) would be
obtained only for a standard efficiency of $\epsilon = 0.1$ which as seen
from Fig.~8 \& 9  does not fit the data point for only 
baryonic matter accretion.

\subsection{BH masses from energy conservation arguments}

A gravitational  system in which  molecular clouds provide the
interaction path between DM particles and feeding the BH leads to a strong
limit on the maximum mass of the BH.
  For an initial distribution of DM in the halo
  without a BH, 
  the
  total energy of the system is given by
\begin{equation}
E_{i}=\frac{-3GM_{DM}^{2}}{5R_{DM}} \, \, \, \, ,
\end{equation}
where $M_{DM}$, $R_{DM}$  stand for the total  mass of the DM
halo and  its size, respectively.
After the BH is formed  at the DM halo center, its total
energy is given by
\begin{equation}
E_{f}=\frac{-3GM_{BH}^{2}}{5R_{s}} \, \, \, ,
\end{equation}
with $R_{s}$ being the Schwarzschild radius
\begin{equation}
R_{s}=\frac{2GM_{BH}}{c^{2}} \, \, \,.
\end{equation}
Equating $E_{f}$ to $E_{i}$ as a limit,  we arrive at the following constraint
\begin{equation}
\frac{M_{DM}}{M_{BH}}=\left(\frac{R_{DM}}{R_{s}}\right)^{1/2} \,\,\,,
\end{equation}
from which
we derive the BH mass limit 
\begin{equation}
M_{BH}^{max}=\frac{2GM_{DM}^{2}}{c^{2}R_{DM}} \,\,\,.
\end{equation}
Assuming an isothermal gas sphere distribution for the DM halo,
i.e. $M_{DM}=2 \sigma^{2}R_{DM}/G$, with $\sigma$ being the DM
velocity dispersion, we arrive at the final expression for
the  BH mass
\begin{equation}
M_{BH}^{max}=\frac{8\sigma^{4}R_{DM}}{c^{2}G} \, \, \, ,
\end{equation}
or
\begin{equation}
M_{BH}^{max}=4\left(\frac{\sigma}{c}\right)^{2}M_{DM} \,\,\,.
\end{equation}
The last two equations constrain the properties of  the DM
matter halo to the BH and could be considered as the theoretical 
interpretation  of the  $M_{BH} \; - \; \sigma$ observed correlation 
 in elliptical galaxies. Here, we note that $\sigma$ stands for the velocity dispersion of DM.
 Using the above equations, the BH mass
limit $M_{BH}^{max}$ is obtained to be $\sim 10^{2}M_{\odot}$,
$\sim 10^{6}M_{\odot}$, and 
$\sim 10^{9}M_{\odot}$ for dwarf galaxies, the Milky Way and M87, respectively.
The corresponding DM halos have masses of $\sim 4\times 10^{9}M_{\odot}$,
$\sim 3\times 10^{12}M_{\odot}$ and $\sim 4\times 10^{14}M_{\odot}$.
In order to get the above estimates for the maximum BH mass, we have used 
a DM
halo size of $R_{DM} \sim 9 \  {\rm kpc}$ for dwarf galaxies, 
$R_{DM}\sim 200 \ {\rm kpc}$ for the Milky Way
and  the size of M87 is assumed to be $ \sim 5 \ {\rm Mpc}$.

\section{Discussion \& conclusion}
In this paper, we have 
investigated the growth of  a stellar seed BH immersed at the
center of DM halos with degenerate FBs of mass from
$\sim 10^{3}M_{\odot}$ to $\sim 10^{6}M_{\odot}$.
Using the Pauli exclusion principle, we have established that the BH accretion rate 
 strongly depends on the mass $m_{f}$ of the 
 fermions as  $\dot{M}_{BH}\sim m_{f}^{4}M_{BH}^{2}$ and thus establish
for the first time the relationship between  BH growth and fermionic DM. 
We have shown that in order to fit the DM distribution in the
Galaxy with such degenerate cores, the  DM particles 
should be in the  $ 12 {\rm keV/c}^{2} \stackrel {\textstyle <}{\sim} m_{f}\stackrel {\textstyle <}{\sim}
450{\rm keV/c}^{2}$ mass range. 
We have shown that such DM masses could be  used to fit the
distribution of DM in dwarf galaxies.
  FBs of masses $10^{3 \; - \; 6} M_{\odot}$
could only exist in such galaxies where the density drops off as  $1/r^{2}$  at large distances.
Dwarf galaxies as well as cluster of galaxies do not host FBs as their data can only be fitted
by a non - degenerate Fermi - Dirac distribution of the King type.
We have argued that the merging of dwarf galaxies
would lead to the formation of galaxies with degenerate FBs.
The wideness of  the fermion mass in the keV range is due to the FB mass range
 from $10^{3}$ to $10^{6} M_{\odot}$
that we have adopted in this paper.
Our main assumption is the  use of fermions as DM candidates in galaxies.
 However, if one uses bosons instead of fermions, the range of DM particle masses
 would of course differ  from the one  obtained in this paper. 
The range of fermions used in our paper
is in conflict with the Lee-Weinberg lower  limit of $\sim 2 {\rm GeV/c^{2}}$
 on the fermion mass (Lee \& Weinberg \cite{weinberg77}). This is due to the fact that the derivation
 of the Lee-Weinberg limit assumes a  freezout from equilibrium distributions. Our simple  model
uses the chemical potential which  allows for non equilibrium distributions and this  
 might modify the Lee-Weinberg argument; this remains to be demonstrated.
     
If we use heavy fermions with  masses of 
about  $1 {\rm GeV/c^{2}}$, then  according to equation (\ref{eq:ov}) the maximum mass allowed for the degenerate Fermi 
core would only be of about $1M_{\odot}$, which is not enough to grow  a stellar mass BH
 to $10^{3}M_{\odot}$ in about $10^{8}$ years.
On the other hand,  very light fermions i.e. $m_{f}   \stackrel {\textstyle <}{\sim} 
12 {\rm keV/c}^{2}$
would generate very massive degenerate FBs of masses greater than $10^{6}M_{\odot}$ and the BH would grow to
$10^{9}M_{\odot}$ in a very short time, i.e $z << 6.41$.

The growth of seed BHs from DM accretion is investigated using the
quantum cascade mechanism
 upon which low angular momentum DM particles at the inner Fermi surface
are first consumed by the BH and then due to a high degeneracy pressure,
higher angular momentum particles are pushed inwards and the process
continues until the entire degenerate FB is consumed by the BH.
Moreover, molecular clouds have been used  as perturbers of 
 DM particle orbits outside the FB
 and we have shown that the BH grows faster than the FB.
 After the BH has consumed  the entire FB, it then grows by Eddington - limited baryonic
  accretion to
higher masses of $\sim 10^{9}M_{\odot}$ at redshifts $z \sim 6.41$.
We also point out that  molecular clouds of mass $10^{10}M_{\odot}$ have also been 
 detected in the host galaxy of the same quasar at a redshift of
$z \sim 6.41$ (Walter et al. \cite{walter03}).

We have also constrained  the possible starting time of
accretion, i.e. the time of BH seed formation. 
From our analysis, the  mass of a $3\times 10^{9}M_{\odot}$ BH  in the 
quasar SDSS J114816.64+525150.3  at a redshift of $z=6.41$ can be fitted 
exactly if the accretion process starts at a time of about  $2 \times
10^{8}$ years, which corresponds to the reionization time.

The seed BH mass is found to be in the range from a few solar masses up to an upper limit
of $\sim 10^{4}M_{\odot}$. For a seed BH mass  of $ 10^{3\, - \, 4}M_{\odot}$, Eddington
baryonic matter accretion would be enough to cause the seed BH to grow into  a
supermassive BH of $3\times 10^{9}M_{\odot}$ mass.
The data point at a redshift of $z=6.41$  can be fitted  by only  
Eddington baryonic matter accretion with an  efficiency of $\epsilon \sim 0.01$.

Our model provides a method to find the DM particles mass. 
If it is found that there is a clear lower mass cut of
$10^{3}$ to $10^{6} M_{\odot}$ in the distribution of BH masses, then 
this  mass can be used for the mass of the FB to find the corresponding mass $m_{f}$ of the 
fermions which will be in the range of 12 keV to 450 keV.
If on the other hand the BH mass distribution is a continuous function, then our model of BH
growth  with DM will probably be ruled out.

The postulated DM particles in this paper were non - relativistic at the decoupling time and 
are usually called cold dark matter
particles (CDM). The latter have to be neutral, stable or quasi-stable and have to weakly interact with ordinary matter.
As mentioned in section 2.1 , these particles could be axions which have been investigated by DAMA/NAI (Bernabei et al.
\cite{bernabei01}). The heavier particles  of mass above $1 {\rm GeV/c^{2}}$ could also be of the class of DM 
candidates named WIMPS (Weakly Interactive 
Massive particles). However, in the standard model of particle physics, CDM cannot be  
suitable candidates for particles.
Thus, a new window beyond the standard model of particle physics has to accommodate these particles 
 for our model of BH growth to work.
The DAMA/NAI experiment (see Bernabei et al. \cite{bernabei05} for a review) which aims at the verification of 
the presence of DM particles in the Galactic halo,
 will be able to confirm whether GeV WIMPS or  axions of masses of 12 to 450 keV could exist in nature.  
  
Observations have shown that the masses of  
supermassive BHs at galactic centers correlate with the masses of the 
host bulges, i.e. $M_{BH} \approx  0.002 \  M_{bulge}$ 
(Haering \& Rix \cite{rix04}). This result is obtained in our model as long as the
Eddington limited accretion dominates the final growth of the BH. This happens for BH 
masses of  more than about $ 10^{5} M_{\odot}$ (Wang, Biermann \& Wandel \cite{wang00}).

Mergers  of dwarf galaxies   as 
well as the spinning of BHs would play an important role in the 
growth of the BHs. The consideration of these two effects will be the 
subject of 
 further investigations.
In addition, it would be of great interest to study the growth of BHs from boson DM particles.
Although it has been shown that bosons could provide a good fit to the rotation curves in dwarf galaxies, it is not 
yet clear whether an analogous mechanism could work in galaxies with a $1/r^{2}$ density fall off.

While  it takes only $8.4\times 10^{8}$ years to grow supermassive BHs in 
most distant quasars, the Galactic center 
might have grown to its current 
mass of $\sim 10^{6}M_{\odot}$ with only DM accretion in  a Hubble time. 
In the following  paper, we will address the growth of the Galactic center BH.

\begin{acknowledgements}
    We thank Yiping Wang for providing us with the data file for baryonic matter 
growth. We are also grateful to Heino Falcke and Rainer Spurzem for 
useful discussions. FM research is supported by 
the Alexander von Humboldt Foundation.
Work with PLB has been supported through the AUGER theory and membership grant
05 CU1ERA/3 through DESY/BMF (Germany). Further support for the work  with  PLB has come from the
DFG, DAAD, Humboldt Foundation (all Germany), grant 2000/06695-0 from FAPESP (Brasil)
through G. Medina-Tanco, a grant from KOSEF (Korea) through H. Kang and D. Ryu, a grant from
ARC (Australia) through R. J. Protheroe, and European INTAS/ Erasmus/ Sokrates/ Phare grants with
partners V. Berezinsky, L. Gergely, M. Ostrowski, K. Petrovay, A. Petrusel, M. V. Rusu and S. Vidrih.
  
\end{acknowledgements}

\end{document}